# Influence of gas flow rate on liquid distribution in trickle-beds using perforated plates as liquid distributors

Juan-David Llamas, François Lesage, Gabriel Wild[1*]

*Laboratoire des Sciences du Génie Chimique, Nancy-Université, CNRS, 1, Rue Grandville, 54000 NANCY, France.*

**Abstract**

Two wire mesh tomography devices and a liquid collector were used to study the influence of the gas flow rate on liquid distribution when fluids distribution on top of the reactor is ensured by a perforated plate. In opposition to most of the studies realized by other authors, conditions in which the gas has a negative impact in liquid distribution were evidenced. Indeed, the obtained results show that the influence of gas flow rate depends on the quality of the initial distribution, as the gas forces the liquid to "respect" the distribution imposed at the top of the reactor. Finally, a comparison between the two measuring techniques shows the limitations of the liquid collector and the improper conclusions to which its use could lead.

## 1. Introduction

Trickle-beds reactors are widely used chemical reactors in which a gas and a liquid flow cocurrently downwards through a fixed bed of solid particles. Along with their advantages in terms of cost and simplicity, trickle-beds reactors have two main handicaps: they are unable to evacuate big amounts of heat and it is difficult to properly distribute the fluid phases over the catalyst particles. While the first of these two problems will mostly restrain the application field of trickle-beds reactors, the second will directly affect its performances. Indeed, a bad liquid distribution will not only lead to an inefficient utilization of catalyst but could also be the cause of hot spot formation.

Among the factors affecting liquid distribution in trickle beds one will find the initial liquid distribution (ensured by the use of a proper distributor), the packing and fluids characteristics, the pre-wetting conditions and the flow rates. Those factors have been widely studied by numerous authors (e.g. [1-10]). Many of their studies however, dealt mostly with small sized columns and/or were performed using liquid collectors; so, in a very recent investigation, Babu *et al.*[10] used a column with 50 mm i.d.. Note that these authors compared collector data with radial spreading data issued from a radial heat transfer model. Although simple to use and quite valuable in many circumstances, the liquid collector can only give access to the macroscopic flow rate distribution at the exit of the reactor. As the flow distribution inside the reactor is not necessarily reflected by the exit distribution, the use of a liquid collector could lead to improper conclusions. For this reason, the present work tries to complete the information obtained with a liquid collector by using two wire-mesh tomography devices that allow studying liquid distribution at two different bed depths. Using two different perforated plates for fluid distribution and operating under different gas and liquid flow rates, the present work focuses on the influence of the gas flow rate on liquid distribution when this kind of distributor is used.

---

[1] Département de Chimie Physique des Réactions, Nancy-Université, CNRS, 1, Rue Grandville, 54000 NANCY, France.
* Corresponding author : E-mail : gabriel.wild@ensic.inpl-nancy.fr

## 2. Experimental setup

*2.1 The reactor*

The experiments were performed using a 0.3 m diameter trickle-bed reactor. The fixed bed, of 1.3 m height, was composed of porous alumina cylinders whose mean length and diameter were 4.3 mm (with a standard deviation of 1.5 mm) and 1.2 mm (with a standard deviation of 0.16 mm) respectively. The internal porosity of the solid particles, measured by helium pycnometry, was 0.46, while the external porosity of the fixed bed, which was sock packed, was 0.47.

Air and water were used as fluids. The superficial velocities considered varied from 0 to 0.34 m/s for the gas and from 0.0016 to 0.0047 m/s for the liquid. All experiments were performed at room temperature (22°C) and pressure (1 bar) and using a pre-wetted bed.

For fluid distribution, two perforated plates were used: perforated plate "A" which had 24 liquid inlets and 4 gas inlets and will be referred to as "uniform distributor" and perforated plate "B" which had 5 liquid inlets and 4 gas inlets and will be referred to as "cross-shaped distributor" (Figure 1). The second distributor is used to mimic an imperfect initial distribution, while the first one is a design actually used in industry a dozen years ago.

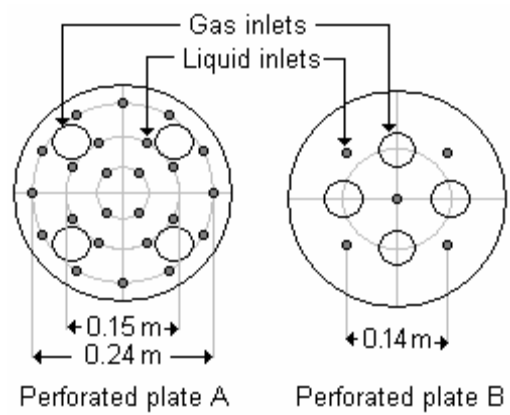

**Figure 1. Perforated plates used for fluid distribution.**

Between the perforated plate and the solid particles, 3 layers of "inert" spherical particles were disposed: a first layer of 3 mm steel spheres that were in contact with the catalyst, a second layer of 5 mm steel spheres over the previous ones and a third layer of 6 mm ceramic beads in contact with the perforated plate. These inert particles occupied about 30 mm of the fixed bed.

Further details on the experimental plant may be found in Llamas [11].

*2.2    The measuring devices*

Two different measuring techniques were used for the experiments: a liquid collector and two wire-mesh tomography devices. The liquid collector, placed at the exit of the reactor, was divided in 9 collecting zones of the same surface (Figure 2) and allowed measuring the macroscopic liquid distribution in terms of flow rate at this point; it is the same as the one proposed by Marcandelli et al.[6]. The quantification of the results obtained with this device was done by means of the maldistribution factor ($M_f$) defined according to Marcandelli et al.[6].

$$M_f = \sqrt{\frac{1}{N(N-1)} \sum_i \left(\frac{Q_i - \overline{Q}}{\overline{Q}}\right)^2} \qquad (I)$$

where $N$ is the number of collecting zones (9 for the present work), $Q_i$ is the flow rate measured in zone i and $\overline{Q}$ is the mean liquid flow rate.

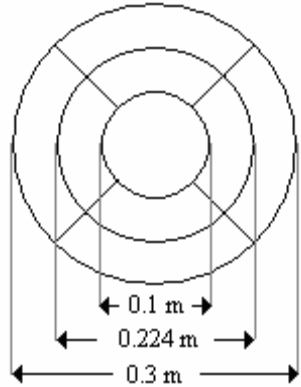

**Figure 2. Liquid collector used for the experiments.**

The wire mesh tomography devices were placed 0.33 m and 0.96 m from the distributor and allowed measuring the liquid distribution in terms of liquid saturation at those two bed depths. More precisely, each wire mesh tomography device is composed of two horizontal planes of parallel wires placed at a short distance between them (4 mm in the present case). The angle between the wires of both planes is 90°, forming thus a square grid where the wires do not touch each other and allowing, by local conductance measurements, to detect the presence of a conductive fluid between them. Each plane contains 19 stainless steel wires of 2 mm diameter, resulting in 313 measurable crossing points per device. The operating principle of wire mesh tomography is simple: if a current is applied to one of the wires of the first (transmitter) plane, it will not be measured on the second (receiver) plane, unless a conductive fluid makes the connection between them. As the conductance of a liquid filament is proportional to its section, for each of the crossing points, the local liquid saturation can be estimated as:

$$\beta_L^i = \frac{\varepsilon_L^i}{\varepsilon} = \frac{C^i}{C^i_{liquid\ filled\ column}} \qquad \text{(II)}$$

Where $C^i$ is the measured conductance for crossing point i and $\beta_L^i$ is the local liquid saturation which is the ratio of the local liquid holdup $\varepsilon_L^i$ to the bed porosity $\varepsilon$. More detailed information about the working conditions, advantages and possible pitfalls of this device can be founded elsewhere [12].

It is important to state that, as the measuring principle of wire mesh tomography reposes on conductance measurements, the use of this technique inside a bed of porous particles is quite complicated. Indeed, once wetted, porous particles remain filled by water and can conduct electricity. In this case, it is thus difficult to make any difference between a set of wetted particles and a real water filament flowing through the reactor. For this reason, between the wires of the tomography device, porous particles were replaced by 2 mm glass beads. However, radial spreading in a bed of glass beads is slow, as shown in [11], and it can safely be assumed that the influence on liquid distribution of a length of 50 mm of glass beads (which corresponds to about 3.8% of the total bed height) is negligible.

### 3. Results

Results were obtained using the two measuring techniques described earlier for two perforated plates and five different liquid superficial velocities ranging from 0.0016 to 0.0047 m/s. For each of the liquid flow rates considered, five gas superficial velocities between 0 and 0.34 m/s were studied.

*3.1 Uniform distributor*

With 24 liquid inlets (equivalent to 340 inlets per m$^2$ of reactor cross-section), perforated plate "A" can be considered as a good liquid distributor. Some of the images obtained using this perforated plate at $U_L$ = 0.0082 m/s are presented in Figure 3. From those images, and the ones obtained for the others gas-liquid superficial velocities combinations, two well know consequences of the gas flow on the reactor hydrodynamics were evidenced. The first one is the logical reduction of the mean liquid saturation as the gas flow rate increases. The second is that, while the entrance of gas inside the reactor modifies deeply the liquid distribution, further increase of its flow rate has only a limited effect on the liquid flow pattern. This is not surprising as in absence of gas flow the mechanisms governing the single phase liquid motion are not the same: with a gravity-driven liquid flow, the distribution is mainly affected by packing topology and capillarity forces.

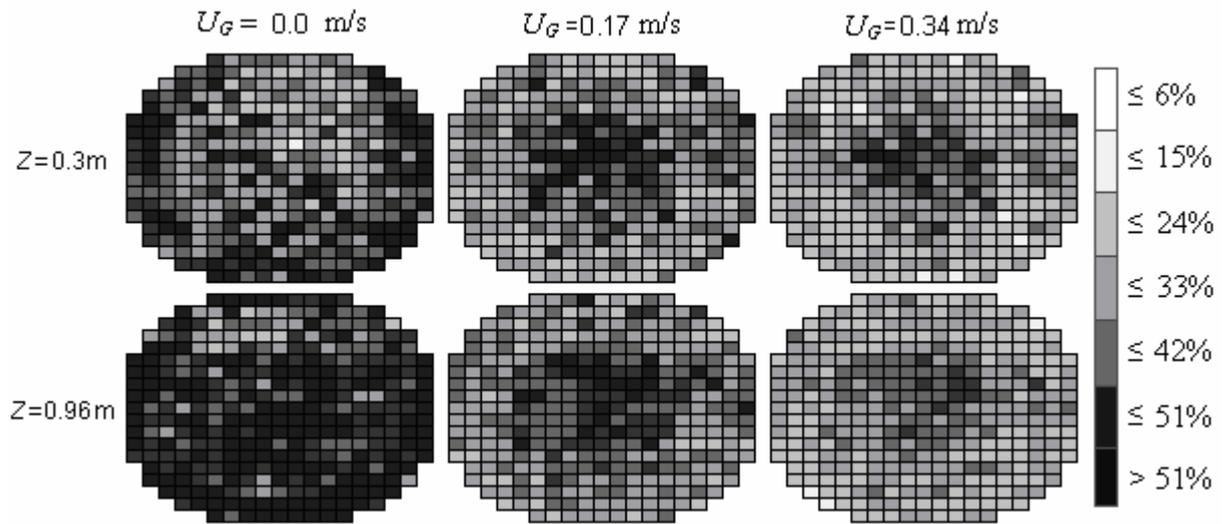

**Figure 3. Uniform distributor: images obtained at $U_L$ = 0.0082 m/s.**

In order to quantify the quality of the liquid distribution, the mean relative standard deviation (*MSD*) is used. The relative standard deviation of a set of measurements is the standard deviation of the measurements divided by the mean value of the same measurements. For each of the two tomography devices, a relative standard deviation is calculated and the average of those two values is called *MSD*. From this point of view, high values of *MSD* are related to bad liquid distributions as it means that local liquid saturation values are not uniform along the fixed bed cross-section.

The obtained results show that, from a liquid saturation point of view and for $U_G \neq 0$, the quality of the liquid distribution does not seem to be severely affected by the gas flow rate (Figure 4). Liquid superficial velocity, however, seems to play an important role as, as it can be seen in Figure 5, the *MSD* value decreases notably when $U_L$ increases. The predominant effect of the liquid flow rate on liquid distribution compared to that of the gas had already been reported, among others, by Sylvester and Pitayagulsarn[1] and Kundu *et al.*[7].

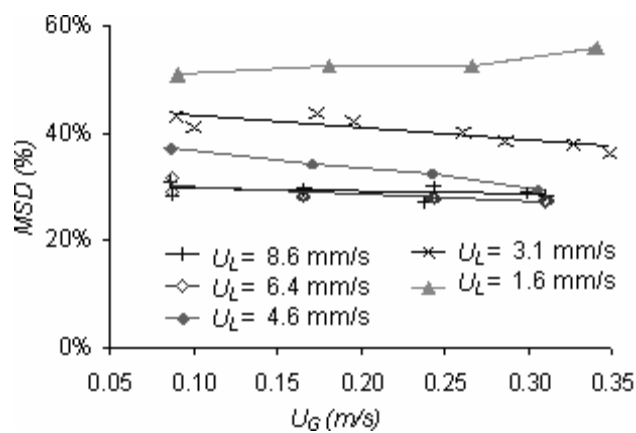

**Figure 4. Influence of $U_G$ on *MSD* for different liquid superficial velocities.**

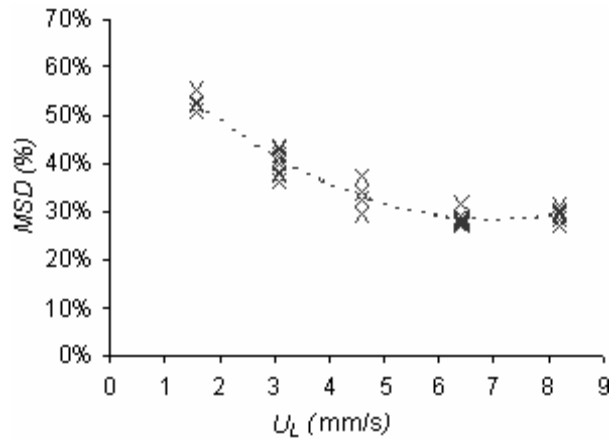

**Figure 5.** Influence of $U_L$ on *MSD* at different values of the gas superficial velocity.

Measurement were also realized with the liquid collector. The maldistribution factor $M_f$ was calculated for all liquid-gas flow rate combinations. Although not of the same intensity for all liquid flow rates, an improvement of the liquid distribution was observed at increasing gas flow rate. The former statement can be easily appreciated from Figure 6, where $M_f$ is plotted against $U_G$ for all the experiments regardless of the liquid flow rate. In this figure, the maldistribution factors for $U_G = 0$ were omitted because their dependence on $U_L$ seems to be stronger. Those $M_f$ values were comprised between 0.15 and 0.26 which shows, as already reported by authors like Marcandelli *et al.* [6] and Møller *et al.* [4], the beneficial effect of having a flowing gas inside the reactor.

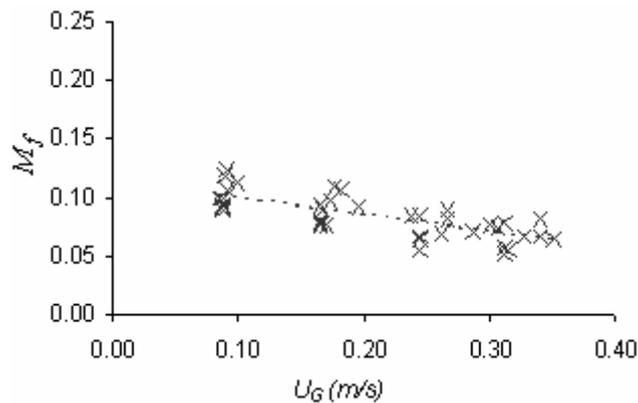

**Figure 6.** Evolution of $M_f$ with gas superficial velocity for all liquid flow rates.

The results obtained with both measuring techniques lead to a first hypothesis about the fluid flow inside the reactor. Indeed, while wire mesh tomography showed that liquid distribution quality is not deeply affected by gas flow, the liquid collector showed that the flow rates at the exit of the reactor gain in uniformity as gas flow rate increases. That could mean that the gas follows preferred paths inside the reactor. The zones of the column concerned by those paths, probably because gas and liquid are injected separately into the column (as perforated plates are used), will have lower liquid saturation than others but, as gas velocity is higher and liquid droplets can be carried along by gas flow, the flow rates at the exit of the reactor will be macroscopically similar for those zones that for zones where liquid saturation is higher but the amount of gas is lower. Flow rates measurements inside the reactor (if possible near the wire mesh tomography devices) would be necessary to get information concerning the validation of the stated hypothesis.

*3.2    Cross-shaped distributor*

With 5 distant liquid inlets, the cross shaped distributor allows getting an insight about the influence of gas flow over individual liquid filaments. Figure 7 shows typical images obtained with this perforated plate. The presented results concern a liquid superficial velocity of 0.0082 m/s at 3 different gas flow rates.

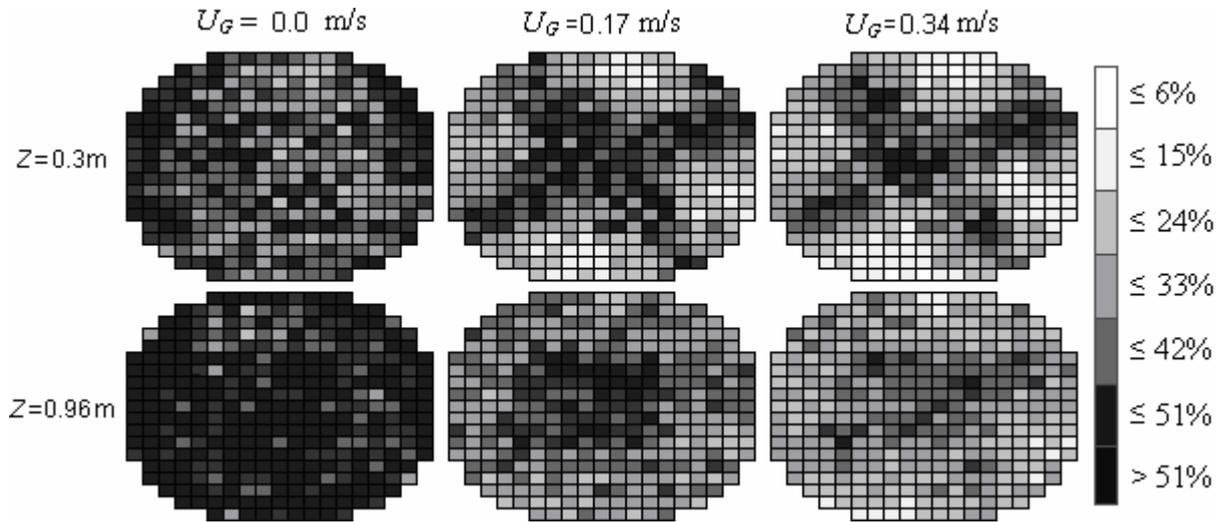

**Figure 7. Cross-shaped distributor: images obtained at $U_L$ = 0.0082 m/s.**

With this set of images, the influence of gas flow rate on liquid distribution becomes clearer. Indeed, when no gas is flowing through the reactor, a liquid flow pattern similar to the one obtained with the uniform distributor is observed: the liquid manages to occupy the whole reactor cross-section before the first tomography device. When the gas flow rate is larger than 0 however, liquid flow pattern changes radically to adopt, at least for a distance superior to one reactor diameter, the flow pattern imposed by the distributor. According to the obtained images, this low quality liquid distribution is emphasized by the increase in gas flow rate.

The effect of gas flow reported previously is more pronounced for high liquid flow rates without completely disappearing for the lowest ones (Figure 8). For the latter, even if the cross-shaped pattern is less evident, the increase of poorly irrigated zones leads to an increase of *MSD* values. It is important to notice that, in order to avoid appreciation errors issued from the increase of liquid saturation with liquid flow rate, the colours assigned to each of the images in Figure 8 are a function of the mean liquid saturation of the concerned image.

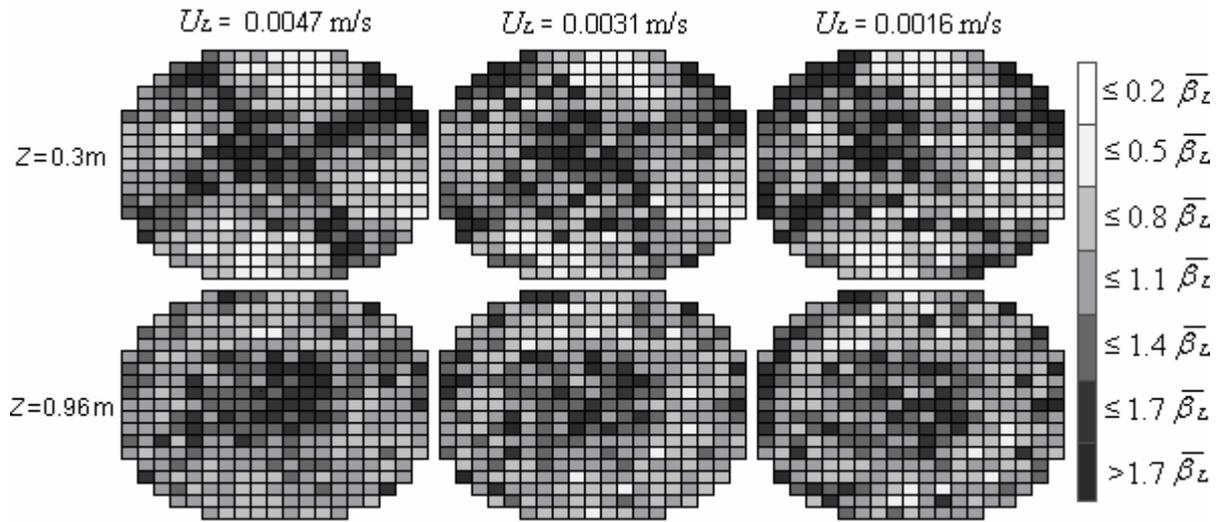

**Figure 8. Cross-shaped distributor: images obtained at $U_G \approx 0.17$ m/s.**

The comparison between the results obtained using the wire mesh tomography and those obtained with the liquid collector is quite interesting. Indeed, as presented in Figure 9 where the maldistribution index ($M_f$) and the *MSD* are both plotted against $U_G$ for $U_L = 0.0082$ m/s, the effect of gas obtained with both techniques is not the same. While the collector shows an improvement in liquid distribution quality with gas in terms of liquid flow rate at the exit of the reactor, the tomography measurements show (as discussed early) a deterioration of liquid distribution quality with the increase of gas flow rate in terms of liquid saturation.

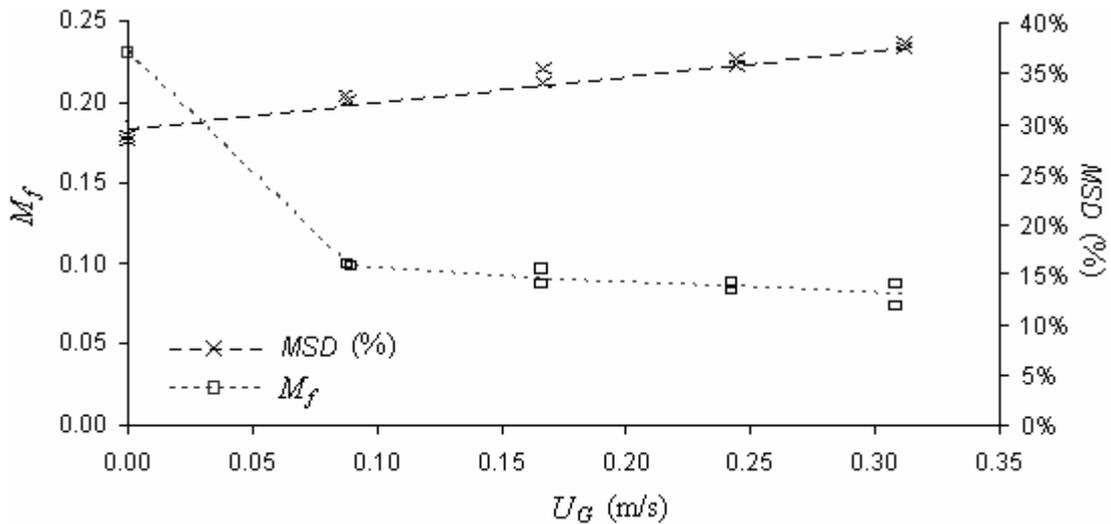

**Figure 9. Comparison between the results obtained with the liquid collector and the wire mesh tomography devices.**

In order to avoid scale problems, local liquid saturation values were averaged to match the 9 measuring zones defined by the liquid collector. The 9 liquid saturation values thus obtained were then used to calculate a $M_f$ kind of parameter (using, evidently, liquid saturation instead of liquid flow rates in equation I) that could be compared to the $M_f$ values measured using the liquid collector. As the $M_f$ values obtained with the wire mesh tomography device were significantly lower than those obtained with the liquid collector, all values were normalized by dividing them by the largest measured value (0.23 for the liquid collector and 0.06 for wire mesh tomography). Results obtained (Figure 10) show a similar behaviour as that evidenced in Figure 9.

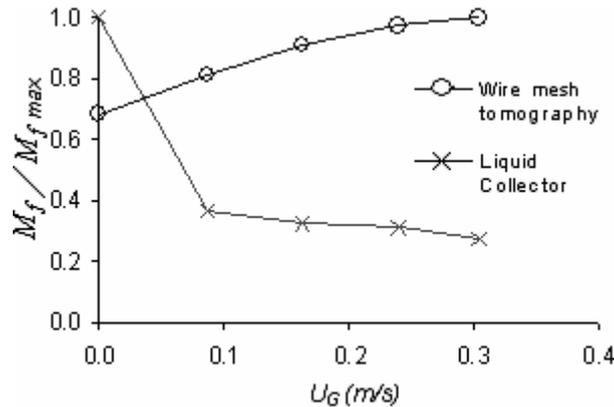

**Figure 10. Maldistribution factors obtained with wire mesh tomography and with the liquid collector at the same scale.**

An explanation of this behaviour is difficult to establish. On the one hand, this could be a consequence of the hypothesis proposed earlier, according to which the gas will follow preferred paths while carrying along liquid droplets (especially when taking a look of the images obtained 0.33 m from the distributor). On the other hand, it is true that liquid distribution seems to improve with the distance from the distributor. Indeed, in Figure 9, the increase in *MSD* with gas flow is mostly the consequence of the increase of the standard deviation measured for the first tomography device, 0.33 m from the distributor. For the second device, the standard deviation does not seem to be altered by gas flow and remains mostly constant. Following this behaviour, it is possible to imagine that the liquid distribution could improve near the exit of the bed to give the results obtained with the collector. Anyhow, and even if the first of the two explanations seems more reasonable, the obtained results do show the limitations of the liquid collector and the effects of gas on liquid distribution.

## 4. Conclusions

The use of two different measuring devices allowed studying the influence of gas flow rate on liquid distribution when perforated plates are used for fluid distribution. When using a good liquid distributor, the results showed that, as stated by authors like Marcandelli *et al.*[6] and Møller *et al.*[4], it is the presence of a flowing gas, more than its flow rate, which has a clear positive impact on the liquid distribution inside the reactor. It was also observed that in general terms, increasing the gas flow rate enhances the quality of the liquid distribution. This kind of behaviour has already been reported, among others, by Weekman and Myers[1], Saroha *et al.*[5], or more recently by Kundu *et al.*[7]. However, in contrast with the results reported by those authors, the measurements showed that the gas can also have a negative impact on liquid distribution. Indeed, when a bad liquid distributor is used (cross-shaped distributor), although the liquid collector shows an improvement in liquid distribution quality, this is not the case of the overall liquid distribution measured in terms of local liquid saturations. According to these results, two main conclusions can be proposed:

- For a distance superior to one reactor diameter, the introduction of gas flow rate causes the liquid to follow the flow pattern imposed by the distributor. If a good liquid distributor is used, the overall influence of the gas will be positive. If a bad liquid distributor is used, however, the presence of the gas will have a negative impact over liquid distribution.

- Although simple to use and quite valuable in many circumstances, the known limitations of the liquid collector were exposed here as it was showed that the sole consideration of exit flow rates can lead to incomplete distribution quality impressions. Indeed, the use of the cross shaped distributor showed that while the overall liquid distribution deteriorates with gas flow (in terms of liquid saturations), the results obtained with the liquid collector showed the opposite behaviour (in terms this time of liquid flow rates).

The presented results show also the importance of considering liquid maldistribution as a function of the quantity that is being measured. It is obviously not the same to consider liquid maldistribution in terms of liquid saturation as liquid maldistribution in terms of liquid flow rates or liquid velocities. Neither the results obtained with the collector nor those obtained with the tomography can be considered as wrong. Those results however, are incomplete if there are considered separately because, at it was expressed before, maldistribution can manifest itself differently depending of the measured quantity.

Finally, it is important to keep in mind that the results presented here concern only reactors in which perforated plates are used for liquid distribution. This consideration could explain up to some extent the differences between the results present here and the behavior reported by other authors. In the case of distributors where gas and liquid are mixed before entering the reactor (e.g. gas chimney trays) the influence of gas flow rate will probably manifest itself in a different way.

**Acknowledgment**

The authors wish to thank "Total Raffinage" not only for financial support, but also for providing the catalyst support particles, and for numerous discussions.**Notation**

$C^i$ = Conductance at crossing point i [Siemens]
$N$ = Number of liquid collecting zones in the liquid collector.
$MSD$ = Mean relative standard deviation.
$M_f$ = Maldistribution factor.
$Q$ = Liquid flow rate [m$^3$/s]
$U_i$ = Superficial liquid velocity of phase i [m/s]
$Z$ = Distance from the distributor [m]
$\beta_L$ = Liquid saturation = $V_L/\varepsilon V_R$
$\varepsilon$ = Mean bed external porosity

**References**

(1) Weekman, V. W. J.; Myers, J. E. (1964). Fluid-flow characteristics of concurrent gas-liquid flow in packed beds. *A.I.Ch.E. Journal* **1964**, *10(6)*, 951.10